\documentclass[manuscript,times]{aastex62}
\graphicspath{{./}{figures/}}

\submitjournal{ApJL}

\shorttitle{Development of Magnetosheath Turbulence}
\shortauthors{Li et al.}

\usepackage{subfigure}
\usepackage{url} 
\usepackage{amsmath}
\usepackage{multirow}
\begin{document}

%
%


\title{Evolution of the Earth's Magnetosheath Turbulence: A statistical study based on MMS observations}

%
%

\correspondingauthor{Hui Li, Wence Jiang}
\email{hli@nssc.ac.cn, jiangwence18@mails.ucas.edu.cn}

\author[0000-0002-4839-4614]{Hui Li}
\affiliation{State Key Laboratory of Space Weather, National Space Science Center, CAS, Beijing, China}
\affiliation{University of Chinese Academy of Sciences, Beijing, China}

\author[0000-0002-8055-4569]{Wence Jiang}
\affiliation{State Key Laboratory of Space Weather, National Space Science Center, CAS, Beijing, China}
\affiliation{University of Chinese Academy of Sciences, Beijing, China}

\author{Chi Wang}
\affiliation{State Key Laboratory of Space Weather, National Space Science Center, CAS, Beijing, China}
\affiliation{University of Chinese Academy of Sciences, Beijing, China}

\author{Daniel Verscharen}
\affiliation{Mullard Space Science Laboratory, University College London, Dorking RH5 6NT, UK}
\affiliation{Space Science Center, University of New Hampshire, Durham, NH 03824, USA}

\author{Chen Zeng}
\affiliation{State Key Laboratory of Space Weather, National Space Science Center, CAS, Beijing, China}
\affiliation{University of Chinese Academy of Sciences, Beijing, China}

\author{C. T. Russell}
\affiliation{University of California, Los Angeles, CA, USA}

\author{B. Giles}
\affiliation{NASA Goddard Space Flight Center, Greenbelt, MD, USA}

\author{J. L. Burch}
\affiliation{Southwest Research Institute, San Antonio, TX, USA}


\begin{abstract}

Composed of shocked solar wind, the Earth's magnetosheath serves as a natural laboratory to study the transition of turbulence from low Alfv{\'e}n Mach number, $M_\mathrm{A}$, to high $M_\mathrm{A}$. The simultaneous observations of magnetic field and plasma moments with unprecedented high temporal resolution provided by NASA's \textit{Magnetospheric Multiscale} Mission enable us to study the magnetosheath turbulence at both magnetohydrodynamics (MHD) and sub-ion scales. Based on 1841 burst-mode segments of MMS-1 from 2015/09 to 2019/06, comprehensive patterns of the spatial evolution of magnetosheath turbulences are obtained: (1) from the sub-solar region to the flanks, $M_\mathrm{A}$ increases from $<$ 1 to $>$ 5. At MHD scales, the spectral indices of the magnetic-field and velocity spectra present a positive and negative correlation with $M_\mathrm{A}$. However, no obvious correlations between the spectral indices and $M_\mathrm{A}$ are found at sub-ion scales. (2) from the bow shock to the magnetopause, the turbulent sonic Mach number, $M_{\mathrm{turb}}$, generally decreases from $>$ 0.4 to $<$ 0.1. All spectra steepen at MHD scales and flatten at sub-ion scales, representing a positive/negative correlations with $M_\mathrm{turb}$. The break frequency increases by 0.1 Hz when approaching the magnetopause for the magnetic-field and velocity spectra, while it remains at 0.3 Hz for the density spectra. (3) In spite of some differences, similar results are found for the quasi-parallel and quasi-perpendicular magnetosheath. In addition, the spatial evolution of magnetosheath turbulence is found to be independent of the upstream solar wind conditions, e.g., the Z-component of the interplanetary magnetic field and the solar wind speed.

\end{abstract}

\section{Introduction}

Characterized by disordered fluctuations over a large range of scales, space plasma turbulence is of great importance in space physics due to its ubiquitous role in converting the fluctuation energy from large scales to small scales and eventually dissipating in collisionless magnetized plasmas \citep{Schekochihin09,Bruno and Carbone 2013}. To measure and study the multiscale nature of turbulence, one of the most common and insightful ways lies in the analysis of the power spectral density (PSD) of the turbulent fluctuations. From that perspective, the PSD of magnetic-field turbulence in the solar wind can generally be characterized by four distinguishable dynamical ranges of scales \citep[e.g.][and the references therein]{Alexandrova09,Sahraoui09,Alexandrova13,Bruno and Carbone 2013,Goldstein15,Huang17}: (1) a scaling of $\sim f^{-1}$ in the energy-containing range; (2) a scaling of $f^{-5/3}$ \citep{Kolmogorov41} or $f^{-3/2}$ \citep{Iroshnikov64,Kraichnan65} in the inertial range or at magnetohydrodynamics (MHD) scales; (3) a scaling of $\sim f^{-x}$ at sub-ion scales with a broader range of slopes, $x \in [-3.1, -2.3]$; (4) even steeper scaling at electron scales.

The Earth's magnetosheath is a highly turbulent region bounded by the bow shock and the magnetopause. For the magnetic energy spectra in the magnetosheath, some similarities with those in the solar wind have been shown in previous studies \citep[e.g.,][]{Alexandrova08a,Riazantseva17,Chhiber18}, for example, the existence of the $\sim f^{-1}$ scaling at large scales and the Kolmogorov spectral index -5/3 at MHD scales \citep{Alexandrova08b,Huang17,Chhiber18}, a broad range of slopes, [-4, -2], at sub-ion scales \citep{Czaykowska01,Alexandrova08b,Safrankova13,Huang14,Chen17,Zhu19}, and steeper spectra at electron scales \citep{Matteini17,Macek18}. However, due to the existence of multiple origins of waves and instabilities \citep{Fairfield76,Omidi94}, magnetosheath turbulence is more complicated than the turbulence in the solar wind. Firstly, the bow shock and the magnetopause influence the magnetosheath turbulence properties \citep{Gurnett79,Rezeau01,Sahraoui06,Rakh18}. Secondly, strong temperature anisotropy generally observed in the magnetosheath can generate various instabilities under different conditions, likely the Alfv\'en/ion-cyclotron instability, the mirror-mode instability, the fast magnetosonic/whistler instability, and the fire-hose instability \citep[e.g.,][]{SK93,QS96,Gary98,HM00,Guicking12,Kunz14,Verscharen16}, which is verified by many studies \citep[e.g,][]{AF93, Anderson94, Schwartz96, Lucek01, Czaykowska01, Sahraoui06, Chen17, Voros19, Teh19, Zhao19a, Zhao19b}. Thirdly, other turbulent fluctuations related to local structures, e.g. current sheets, magnetic islands and vortices, can further complicate the magnetosheath turbulence picture \citep[e.g.,][]{Alexandrova08a, Karimabadi14, Huang18}. In addition, the turbulence properties are also different in the quasi-parallel and quasi-perpendicular magnetosheath \citep[e.g.,][]{Czaykowska01,Shevyrev07,Macek15,Breuillard18,Rakh18b}. Here, the quasi-parallel magnetosheath is defined as the magnetosheath behind a bow shock with quasi-parallel field geometry and the quasi-perpendicular magnetosheath likewise.

In the Earth's magnetosheath, the solar wind is subsonic after crossing the bow shock, although it returns to supersonic in the flanks. Thus, the magnetosheath is a good environment to investigate the evolution of turbulence from a small Alfv{\'e}n Mach number, $M_\mathrm{A}$, to a large $M_\mathrm{A}$. However, the spatial evolution of magnetosheath turbulence has not been comprehensively addressed except for a few studies. Based on a previous case study, the intermittency of plasma turbulence increases in amplitude and anisotropy away from the bow shock \citep{Yordanova08}, and the break frequency of ion-flux spectra evolves to higher frequencies approaching the magnetopause \citep{Rakh17}. From a statistical perspective, \citet{Guicking12} finds a decay of wave intensity of low-frequency magnetic-field fluctuations along the streamlines in the Earth's magnetosheath, which quantitatively agrees with the theoretical concept of freely evolving/decaying turbulence; \citet{Huang17} finds that the $\sim f^{-1}$ spectral scaling of magnetic-field spectra at MHD scales is more likely located in the vicinity of the bow shock, while the Kolmogorov-like scaling at MHD scales located away from the bow shock towards the flank and magnetopause regions. Moreover, the spectral scaling at sub-ion scales flattens from the bow shock to the flank and the magnetopause. Similar results are obtained for the ion-flux spectra in the dayside magnetosheath \citep{Rakh18,Rakh18b}.

Due to the limitations of the time-resolution of plasma instruments, the spatial evolution of turbulence spectral properties in the magnetosheath is rarely addressed simultaneously for the magnetic field, ion density and velocity fluctuations. NASA's \textit{Magnetospheric Multiscale} (MMS) mission measures the magnetic field and plasma moments with unprecedented high-time resolutions in the magnetosheath \citep{Burch16}, which provides a unique opportunity to perform a statistical study on the turbulence evolution in the magnetosheath both at MHD scales and at sub-ion scales. Based on 1841 burst-mode segments of MMS-1 observations from 2015/09 to 2019/06, we perform a comprehensive study of turbulent spectral indices in the dayside terrestrial magnetosheath from MHD scales to sub-ion scales in this study. We focus on the evolution of the spectral indices and break frequencies in the magnetosheath from sub-solar region to the flanks, and from the bow shock vicinity to the magnetopause. In addition, we give a comparison of the quasi-perpendicular and quasi-parallel magnetosheath.


\section{Data Set and Methodology}

The high temporal-resolution data provided by MMS mission enable us to probe magnetosheath fluctuations from MHD scales to sub-ion scales \citep{Pollock16}. The magnetic field is measured by the Fluxgate Magnetometer (FGM) \citep{Russell16}, and the particle moments are obtained by the Fast Plasma Investigation (FPI) instrument. For burst-mode, the time resolutions of FGM and FPI (for ions) are 1/128 s and 0.15 s, respectively. We conduct a statistical survey of spectral parameters of magnetosheath turbulences based on 1841 segments of burst-mode data from the MMS-1 spacecraft from 2015/09 to 2019/06, with the magnetic-field, density, and velocity fluctuations simultaneously investigated both at MHD and at sub-ion scales.

To study the spectral parameters of magnetosheath turbulence, we first obtain the omnidirectional magnetic-field spectrum, the ion density spectrum, and the omnidirectional ion velocity spectrum by performing a Fast Fourier Transform (FFT) on each burst-mode data segment. To ensure that the power spectral density (PSD) is not contaminated by measurement uncertainties, we defind a threshold for the signal to noise ratio ($SN$ in dB) as 10 \citep{Sahraoui13}. For the FGM measurement in the magnetosheath, this constraint is generally satisfied because the $SN$ is always greater than 10 for all the frequencies \citep{Chhiber18}. For the particle moments, we calculate the $SN$ from the FPI data products as

\begin{equation}
\begin{aligned} 
SN = 10 \log_{10}\left[ \frac{\delta X^2}{\delta X^2_{\mathrm{sens}}} \right]
\end{aligned}
\end{equation}
where $\delta X $ and $ \delta X_{\mathrm{sens}} $ are the amplitude of the fluctuation and the level of the sensitivity floor at the spacecraft-frame frequency, $f_{\mathrm{sc}}$.

For single-spacecraft measurements, Taylor's hypothesis is widely used to convert observed timescales to length scales, assuming fluctuations cross the spacecraft at a velocity faster than the dynamical timescale of interest. Taylor's hypothesis is usually satisfied in the solar wind, while in the magnetosheath it could be broken under two major conditions \citep{Howes14, Klein14}: (1) a slow flow with $V_{\mathrm{sw}}/V_\mathrm{A} < 0.3$, where $V_{\mathrm{sw}}$ and $V_{\mathrm{A}}$ are the solar wind speed and Alfv\'en speed, respectively; (2) a dispersive regime in which whistler turbulence dominates. Based on our statistical survey, the slowest flow is found downstream of the bow shock nose as expected, and $V_{\mathrm{sw}}/V_\mathrm{A}$ is typically greater than 0.6. In addition, the dispersive whistler wave is generally beyond the scales discussed here. Thus, Taylor's hypothesis is satisfied in our study.

The PSDs of physical parameters in the magnetosheath usually steepen at kinetic scales, which correspond to the ion characteristic scales, e.g., $f_{\rho \mathrm{i}}$ and $f_{\mathrm{di}}$ \citep{Galtier06, Schekochihin09, Chen14, Safrankova15}. $f_{\rho \mathrm{i}}$, $= V_{\mathrm{sw}}/2\pi \rho_\mathrm{i}$, and $f_{\mathrm{di}}$, $=V_{\mathrm{sw}}/2\pi d_\mathrm{i}$, are the Doppler-shifted frequencies corresponding to the proton gyro-radius, $\rho_\mathrm{i}=V_{\mathrm{th}\perp }/\Omega_\mathrm{i}$, and the proton inertial length, $d_\mathrm{i}=V_\mathrm{A}/\Omega_\mathrm{i}$, respectively. Here, $V_{\mathrm{th} \perp} = \sqrt{2k_\mathrm{B} T_{\mathrm{i}\perp}/m_\mathrm{i}}$ is the perpendicular thermal speed, $\Omega_\mathrm{i}$ is the gyro-frequency, $k_\mathrm{B}$ is the Boltzmann constant, $T_{\mathrm{i}}$ is the proton temperature, $m_\mathrm{i}$ is the mass of a proton, and the subscript $\perp$ denotes perpendicular to the background magnetic field. For simplicity, we automatically determine the break frequency, $f_\mathrm{b}$, by minimizing the chi-square value of a two-stage power-law fitting procedure on the PSDs. Several spectral shapes at the scales of transition from MHD to kinetic regimes have been proposed, such as bumps or plateaus \citep{Riazantseva17,Rakh18b}. To reduce interferences, we obtain the spectral indices at MHD scales ($\alpha_1$) and at sub-ion scales ($\alpha_2$) from the linear least-squares fitting of PSDs over two separated frequency bands. In order to statistically guarantee a reliable scale separation, we set the upper-limit frequency of the MHD range to $f_{\rho \mathrm{i}}$ \citep{Bale05, Chhiber18}, and the lower-limit frequency of sub-ion scales to 0.8 Hz (the mean of $f_{\mathrm{di}}$ plus one standard deviation). Thus, we choose the frequency band as [0.02 Hz, $f_{\rho \mathrm{i}}$] for MHD scales, and we set the frequency band in the sub-ion range to [0.8, 5.0] Hz for magnetic-field spectra and [0.8, 2.0] Hz for density and velocity spectra.  

\begin{figure*}[!htbp]
\centering
\includegraphics[width=38pc]{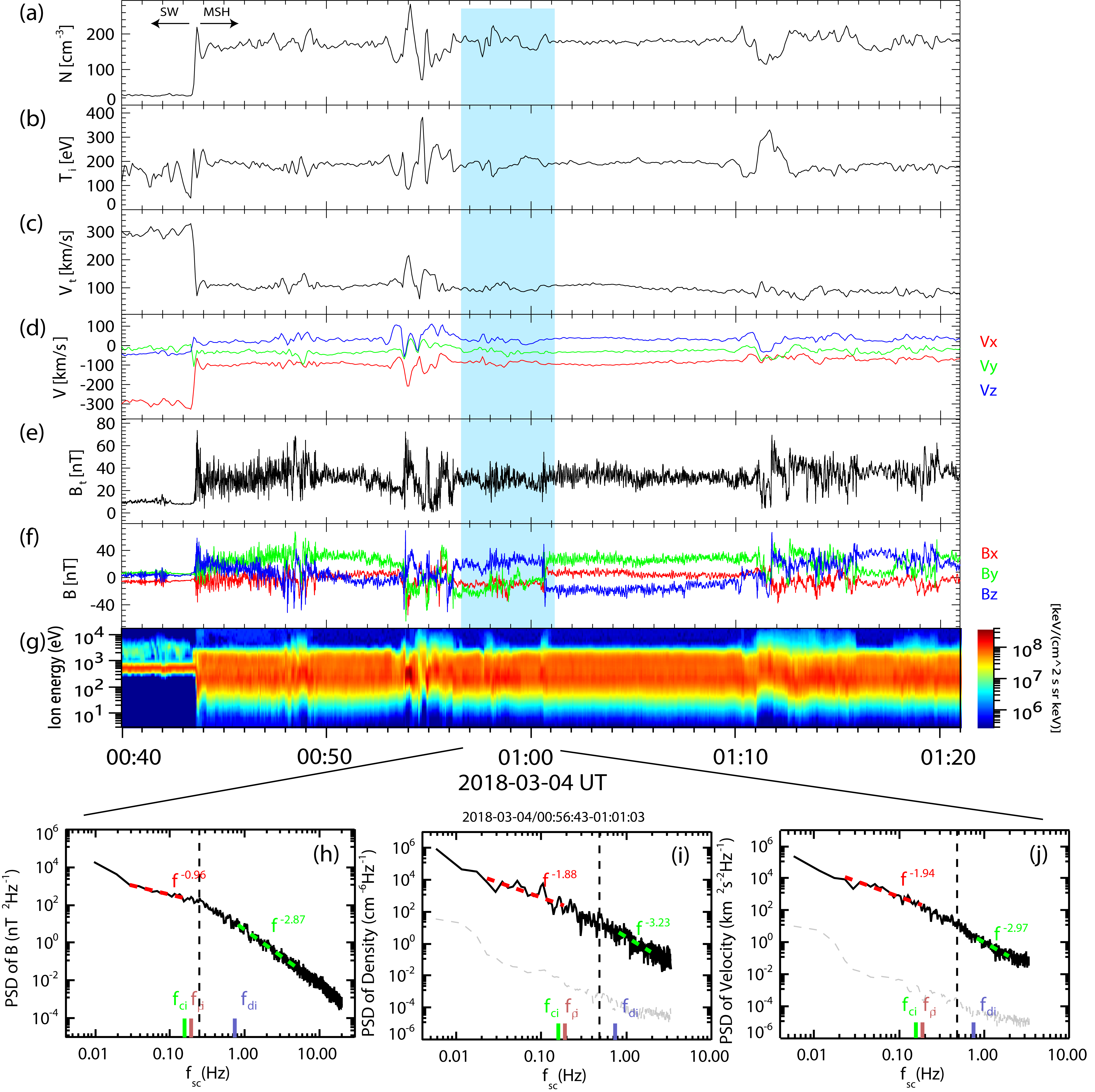}
\caption{An example of magnetosheath observations made by MMS-1 on 2018 March 4. From top to bottom, the panels (a)-(g) show the survey-mode time series: (a) ion number density; (b) ion temperature; (c) the bulk speed of ions; (d) three components of ion velocity in GSE coordinates; (e) magnetic field strength; (f) three components of magnetic field in GSE coordinate; (g) ion energy spectra. The light blue area represents the burst-mode interval in the magnetosheath. Based on the burst-mode data, the omnidirectional magnetic-field spectrum, the ion density spectrum, and the omnidirectional ion velocity are calculated and shown in (h), (i), and (j), respectively. The dashed lines represent the fitting results at MHD scales (in red) and at sub-ion scales (in green). The vertical lines denote different frequencies, such as, the break frequency $f_\mathrm{b}$ (in black), the ion cyclotron frequency $f_{\mathrm{c}\mathrm{i}}$ (in green), the Doppler-shifted frequencies corresponding to the proton gyro-radius $f_{\rho \mathrm{i}}$, and the Doppler-shifted frequencies corresponding to the ion inertial length $f_{\mathrm{d}\mathrm{i}}$. The gray dashed lines in (i) and (j) represent the noise floor of the instrument.}
\label{fig1}
\end{figure*}

Figure \ref{fig1} shows an example of magnetosheath observations made by MMS-1 on 2018 March 4. MMS-1 enters into the magnetosheath (MSH) from the solar wind (SW) at around 00:43 UT and stays in the magnetosheath for more than 30 minutes. During 00:56:43-01:01:03 UT, the burst-mode data denoted by the light blue area are available. In this interval, MMS-1 is located in the sub-solar region (9.8, -2.7, -3.1 $R_\mathrm{E}$ in GSE coordinates) with a local time of about 11:00, and the angle between the bow-shock normal and the interplanetary magnetic-field vector, $\Theta_{\mathrm{Bn}}$ is 46.2$^{\circ}$. The plasma beta value, $\beta_\mathrm{i}$, is about 17.7, the ion temperature anisotropy is about 1.14 ($T_{\mathrm{i}\perp}/T_{\mathrm{i}\parallel}$, where $T_{\mathrm{i}\perp}$ and $T_{\mathrm{i}\parallel}$ are the ion temperature perpendicular to and parallel to the background magnetic field), and the the Alfv{\'e}nic Mach number ($M_\mathrm{A}$) is about 1.99. $\rho_\mathrm{i}$ and $d_\mathrm{i}$ are 84.4 km and 21.0 km, respectively. Thus, $f_{\rho \mathrm{i}}$ and $f_{\mathrm{di}}$ are 0.19 Hz and 0.74 Hz, respectively. 

Based on the burst-mode data, we calculate the omnidirectional magnetic-field spectrum, the ion density spectrum, and the omnidirectional ion velocity and show them in Figure \ref{fig1} (h), (i), and (j), respectively. It is clear that the $SN$ for density and velocity spectra are always greater than 10. All three spectra present well-defined two-stage power laws at MHD scales and at sub-ion scales. For the magnetic-field spectrum shown in Figure \ref{fig1} (h), the spectrum breaks at $\sim$0.25 Hz, near $f_{\rho \mathrm{i}}$. At MHD scales, the spectrum scales as $\sim f^{-0.96}$, shallower than the $f^{-5/3}$ \citep{Kolmogorov41} or $f^{-3/2}$ \citep{Iroshnikov64,Kraichnan65} predictions from MHD, which is consistent with previous results \citep{Czaykowska01, Huang17, Macek18}. The $\sim f^{-1}$ spectral scaling is typically interpreted as the result of the forcing (or energy injection), which indicates the presence of newly-generated local fluctuations at the vicinity of the bow shock. The spectrum steepens to $f^{-2.87}$ at sub-ion scales, a slope consistent with many previous observations in the magnetosheath \citep{Huang14, Macek18}. The formation of a Kolmogorov-like spectrum in the inertial range requires sufficient time (comparable to the nonlinear time) to develop \citep{Chhiber18}. However, the transit time from the bow shock to the MMS-1 location is too short for an inertial range to fully develop, due to the close proximity to the bow shock. In the density spectrum shown in Figure \ref{fig1} (i), the spectrum breaks at $\sim$0.5 Hz. A significant spectral steepening from $f^{-1.88}$ at MHD scales to $f^{-3.23}$ at sub-ion scales is observed. As shown in Figure \ref{fig1} (j), a clear transition of the velocity spectrum from $f^{-1.94}$ at MHD scales to $f^{-2.97}$ at sub-ion scales is observed, with the $f_\mathrm{b}$ $\sim$ 0.5 Hz. At MHD scales, the $f^{-1.94}$ scaling agrees with the prediction ($f^{-2}$) of compressible hydrodynamic Burgers turbulence \citep{GK93} and numerical results \citep{Kim05}. At sub-ion scales, \citet{Safrankova16} show that the velocity spectral indices vary from -5 to -2 and find exceptionally flat spectra ($f^{-2.5}$) when the plasma beta is either very high or very low. For this case, the velocity spectrum is relatively flat, which might be due to a large $\beta_\mathrm{i} \approx 17.7$.

\begin{figure*}[!htbp]
\centering
\includegraphics[width=39pc]{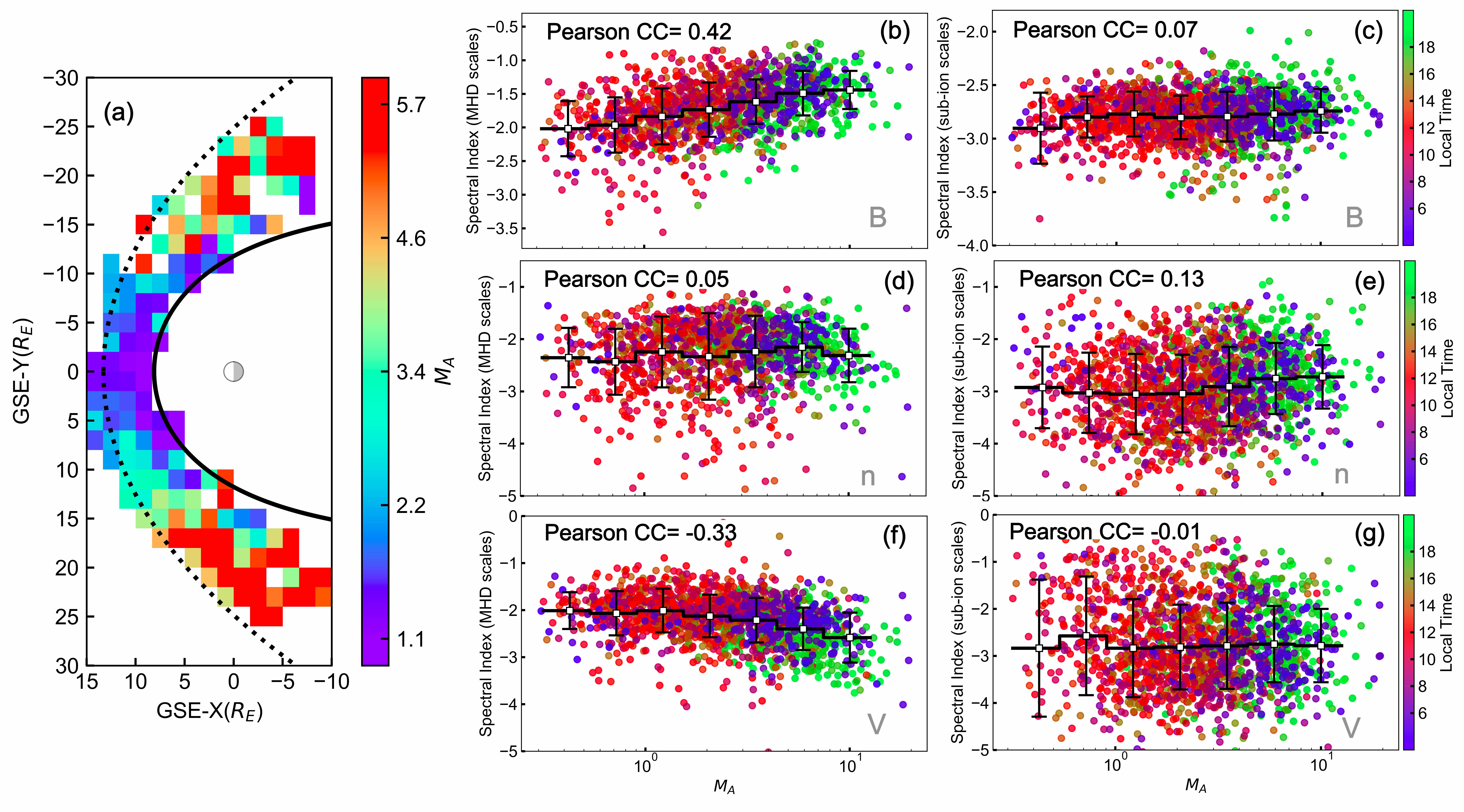}
\caption{Evolution of magnetosheath turbulences from the sub-solar region to the flanks. Panel (a) gives the 2D distribution of the Alfv\'en Mach number, $M_{\mathrm{A}}$, in the GSE-XY plane of the magnetosheath. The black solid and the black dotted curves represent nominal position of the Earth's magnetopause and the bow shock estimated from the models proposed by \citet{Shue97} and \citet{Chao02}, respectively. Panels (b), (d), (f) present the magnetic-field spectral indices, the density spectral indices, and the velocity spectral indices at MHD scales as a function of $M_{\mathrm{A}}$. The results for sub-ion scales are accordingly given in panels (c), (e), (g). The color of the data points indicates the local time. The black horizontal line represents the mean value for each bin, and the vertical line represents the standard deviation. CC represents the correlation coefficient between the spectral slope and $\log_{10} M_{\mathrm{A}}$.}
\label{fig2}
\end{figure*}

\section{Evolution of the Earth's Magnetosheath Turbulence}

\subsection{From the sub-solar region to the flanks}

Figure \ref{fig2} (a) gives the 2D distribution of $M_{\mathrm{A}}$ in the GSE-XY plane of the magnetosheath. The nominal positions of the Earth's magnetopause and the bow shock are estimated from the models proposed by \citet{Shue97} and \citet{Chao02}, respectively. As expected, $M_{\mathrm{A}}$ increases gradually as the solar wind approaches the flanks. In the sub-solar region, the mean $M_{\mathrm{A}}$ is about 1.6, while it grows to 5.2 in the flanks around X = 0 $R_\mathrm{E}$. $M_{\mathrm{A}}$ in the central magnetosheath is generally greater than that in the vicinity of the bow shock and the magnetopause. From top to bottom, the middle three panels present the spectral indices of the magnetic-field, density and velocity spectra at MHD scales as a function of $M_{\mathrm{A}}$. The color of the data points indicates the local time (LT). Evidently, we find a well-organized pattern of $M_{\mathrm{A}}$ vs. LT, in agreement with the 2D distribution of $M_{\mathrm{A}}$ shown in Figure \ref{fig2} (a). For $M_{\mathrm{A}} > 3$, most of the cases are located in the dawn and dusk flanks with LT $\leq$ 8 and LT $\geq$ 16. For $M_{\mathrm{A}} < 2$, most of the cases are located in the sub-solar region with 11 $\leq$ LT $\leq$ 13. For the magnetic-field spectra at MHD scales, the spectral index is positively correlated with $M_{\mathrm{A}}$, with a correlation coefficient (CC) = 0.42. Its mean value changes from -2 for $M_{\mathrm{A}} < 0.8$ to -3/2 for $M_{\mathrm{A}} > 5$. For the density spectra at MHD scales, the spectral index remains $\sim$ -7/3 for different $M_{\mathrm{A}}$ with CC = 0.05, and no obvious relation between spectral index and $M_{\mathrm{A}}$ is found. For the velocity spectra at MHD scales, the spectral index is negatively correlated with $M_{\mathrm{A}}$, with CC = -0.33. Its mean value changes from -2 for $M_{\mathrm{A}} < 0.8$ to -5/2 for $M_{\mathrm{A}} > 5$. The right three panels accordingly give the results for sub-ion scales. No clear relations between the spectral index and $M_{\mathrm{A}}$ are found for all the three types of spectra.

From the sub-solar region to the flanks, the mean spectral indices of the magnetic field change from -2 for $M_{\mathrm{A}} < 0.8$ to -5/3 or -3/2 for $M_{\mathrm{A}} > 5$, implying that the magnetosheath turbulence evolves to be more fully-developed in the flanks. \citet{Alexandrova08a,Alexandrova08b} suggest that the Kolmogorov-like magnetic-field spectrum at MHD scales can be observed in the flanks, where the transit time is long enough for turbulence development. Recently, \citet{Huang17} also find that the $f^{-5/3}$ (Kolmogorov-like) magnetic-field spectra in the frequency range $\sim[10^{-4}, 10^{-1}]$ Hz are only observed away from the bow shock to the flank and magnetopause, which is consist with our results. Here, we present a clear picture of this spatial evolution of MHD turbulence in the magnetosheath and, more importantly, the inverse dependences of magnetic-field and ion-velocity spectral indices on $M_{\mathrm{A}}$.

\subsection{From the bow shock to the magnetopause}

To study the radial evolution of turbulent spectra inside the magnetosheath, the fractional distance between the spacecraft and the magnetopause ($D_{\mathrm{frac}}$) is calculated as proposed by \citep{Verigin06}

\begin{equation}
\begin{aligned} 
D_{\mathrm{frac}}=\frac{r_{\mathrm{sc}}-r_{\mathrm{mp}} }{ r_{\mathrm{bs}} - r_{\mathrm{mp}}}
\end{aligned}
\end{equation}

where $r_{\mathrm{sc}}$, $r_{\mathrm{mp}}$ and $r_{\mathrm{bs}}$ are the radial distances of the spacecraft, the Earth's magnetopause and the bow shock away from the Earth center. $r_{\mathrm{sc}}$ is obtained from the MMS-1 observation directly. $r_{\mathrm{mp}}$ and $r_{\mathrm{bs}}$ can be derived from the empirical models \citep{Shue97, Chao02} with given upstream solar wind conditions provided by the OMNI data. We also perform a comparison of the quasi-parallel and quasi-perpendicular magnetosheath here. Based on the bow shock model \citep{Chao02} and the OMNI data of upstream solar wind with a 5-min temporal resolution, we calculate the angle $\Theta_{\mathrm{Bn}}$ between the interplanetary magnetic field (IMF) and the normal of the bow shock surface corresponding to each data segment. Note that, the location of MMS-1 is projected to the bow shock along the Earth's radial direction. To reduce the uncertainties, we define the quasi-parallel and quasi-perpendicular magnetosheath as $\Theta_{\mathrm{Bn}} < 30^{\circ}$ and $\Theta_{\mathrm{Bn}} > 60^{\circ}$, respectively.

\begin{figure*}[!htbp]
\centering
\includegraphics[width=39pc]{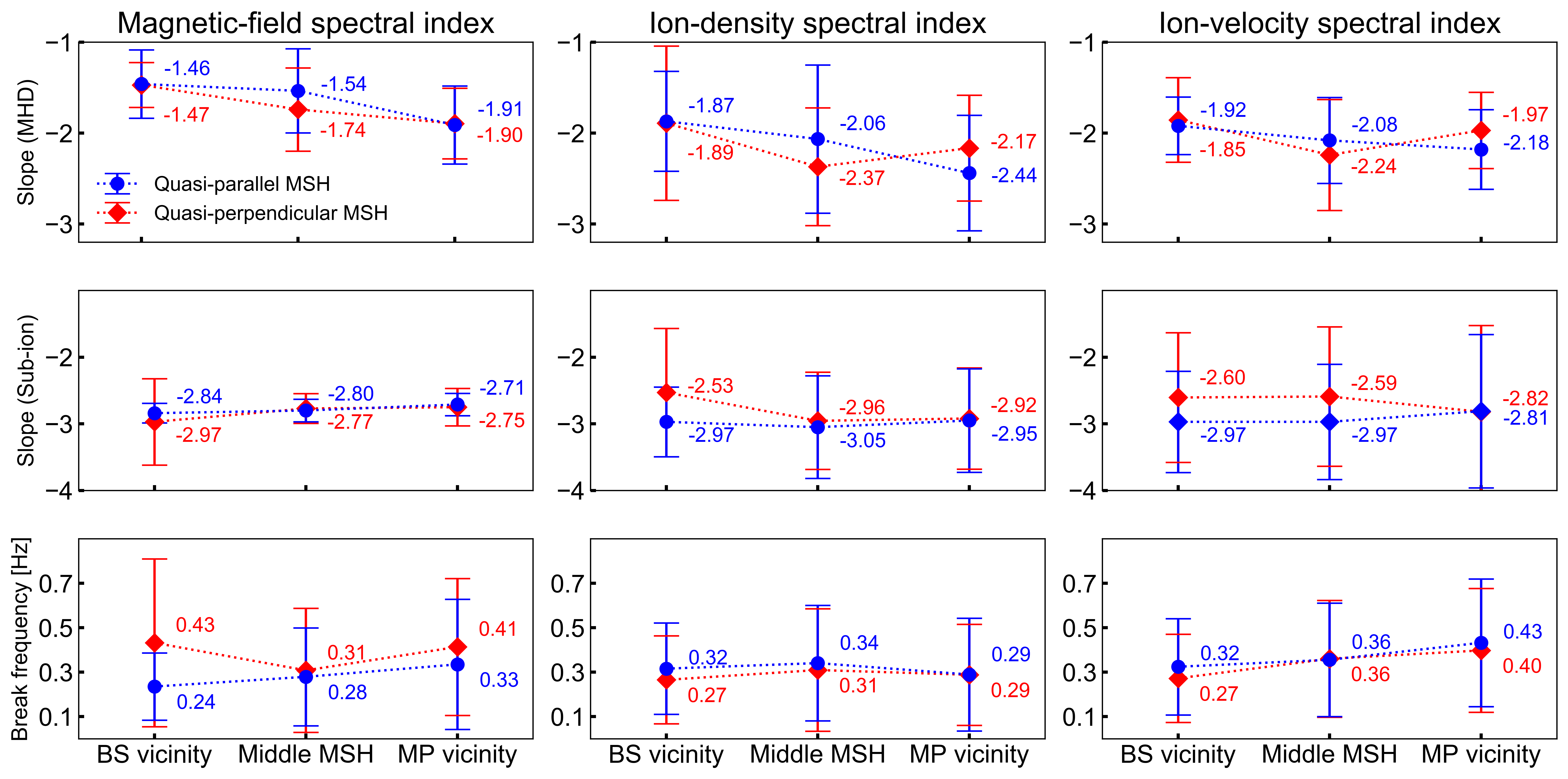}
\caption{Evolution of magnetosheath turbulence from the bow shock to the magnetopause. In the top row, the three panels show the spectral slopes at MHD scales for magnetic-field spectra, density spectra, and velocity spectra, respectively. The results at sub-ion scales are given in the three panels in the middle row. The break frequency is given in the bottom three panels. The averaged values are denoted by blue circles (quasi-parallel magnetosheath) and red diamonds (quasi-perpendicular magnetosheath). The error bars indicate the standard deviation.}
\label{fig3}
\end{figure*}

We show the evolution of magnetosheath turbulence from the bow shock to the magnetopause in Figure \ref{fig3}. For the quasi-parallel magnetosheath, the magnetic-field, density and velocity spectra at MHD scales systematically and monotonically steepens from the bow shock to the magnetopause. The mean value of spectral index for magnetic-field spectra changes from -1.46 in the vicinity of the bow shock to -1.94 in the vicinity of the magnetopause. For density and velocity spectra, it changes from -1.87 to -2.44 and from -1.92 to -2.18, respectively. At sub-ion scales, in-distinctive opposite trends of radial evolution are observed. For the magnetic-field spectra, the spectral slope changes from -2.84 in the vicinity of the bow shock to -2.71 in the vicinity of the magnetopause. For the density and velocity spectra, the spectral index changes from -2.97 to -2.95 and from -2.97 to -2.81, respectively. Compared to the relative large standard deviation, these differences are not significant. The break frequency for magnetic-field and velocity spectra increases when approaching the magnetopause, from 0.24 to 0.33 Hz and from 0.32 to 0.43 Hz, respectively. For the density spectra, it remains around 0.3 Hz. For the quasi-perpendicular magnetosheath, the results are overall similar, except for two differences: (1) For the density and velocity spectra, the steepest spectra at MHD scales occur in the central magnetosheath but not at the magnetopause, and the spectral slope in the bow shock vicinity is flatter than that at the magnetopause; (2) For the magnetic-field spectral, the break frequency is overall greater than that in the quasi-parallel magnetosheath. The lowest value, 0.31 Hz, occurs in the central magnetosheath. In the vicinity of the bow shock and the magnetopause, no clear difference is found, around 0.41 Hz.

At MHD scales, the magnetic-field, density and velocity spectra generally steepen as the spacecraft approaches the magnetopause, especially for the quasi-parallel magnetosheath. Similar findings are reported for the magnetic-field spectra \citep{Czaykowska01, Shevyrev06, Sahraoui06} and the ion-flux spectra \citep{Shevyrev06}, but not for the ion density and velocity spectra. At sub-ion scales, the magnetic-field spectra flatten from the bow shock to the magnetopause, which is consistent with the result of \citet{Rezeau99}. The average spectral index is -2.79 in the quasi-parallel magnetosheath and -2.78 in the quasi-perpendicular magnetosheath, respectively, which coincides with the result of a previous case study \citep{Breuillard18} and agrees with the prediction of kinetic Alfv\'en waves and whistler turbulence models. \citet{Rakh18b} show that the kinetic-scale ion-flux spectra have steeper indices in the bow shock vicinity than near the magnetopause, from -3.2 to -2.8 in the quasi-perpendicular and from -3.4 to -3.0 in the quasi-parallel magnetosheath with an averaged standard deviation of 0.4$\sim$0.5. However, no significant differences are found from the bow shock to the magnetopause considering the large standard deviations in our study. One possibility is that \citet{Rakh18b} focus on the ion-flux, while the density and velocity are investigated separately in this study. Besides, the different numbers of cases potentially contribute as well. Compared to \citet{Rakh18b}, fewer cases (59 vs. 174) in the vicinity of the bow shock and more cases (376 vs. 237) in the vicinity of the magnetopause are used here. Moreover, \citet{Rakh17} show that the break frequency of ion-flux spectra increases from 0.25 Hz in the bow shock vicinity to 0.4 Hz in the magnetopause vicinity, which is similar to our result of velocity spectra, from about 0.3 to about 0.4 Hz. Note that, these values are much less than 0.6-0.8 Hz for ion-flux spectra as shown by \citet{Rakh18b}. This discrepancy may result from the use of different physical parameters (velocity vs. ion-flux) and the effects of possible complex spectral shapes (e.g., bump or plateau found by \citet{Riazantseva17,Rakh18b}) when determining the spectral break.

\begin{figure*}[!htbp]
\centering
\includegraphics[width=38pc]{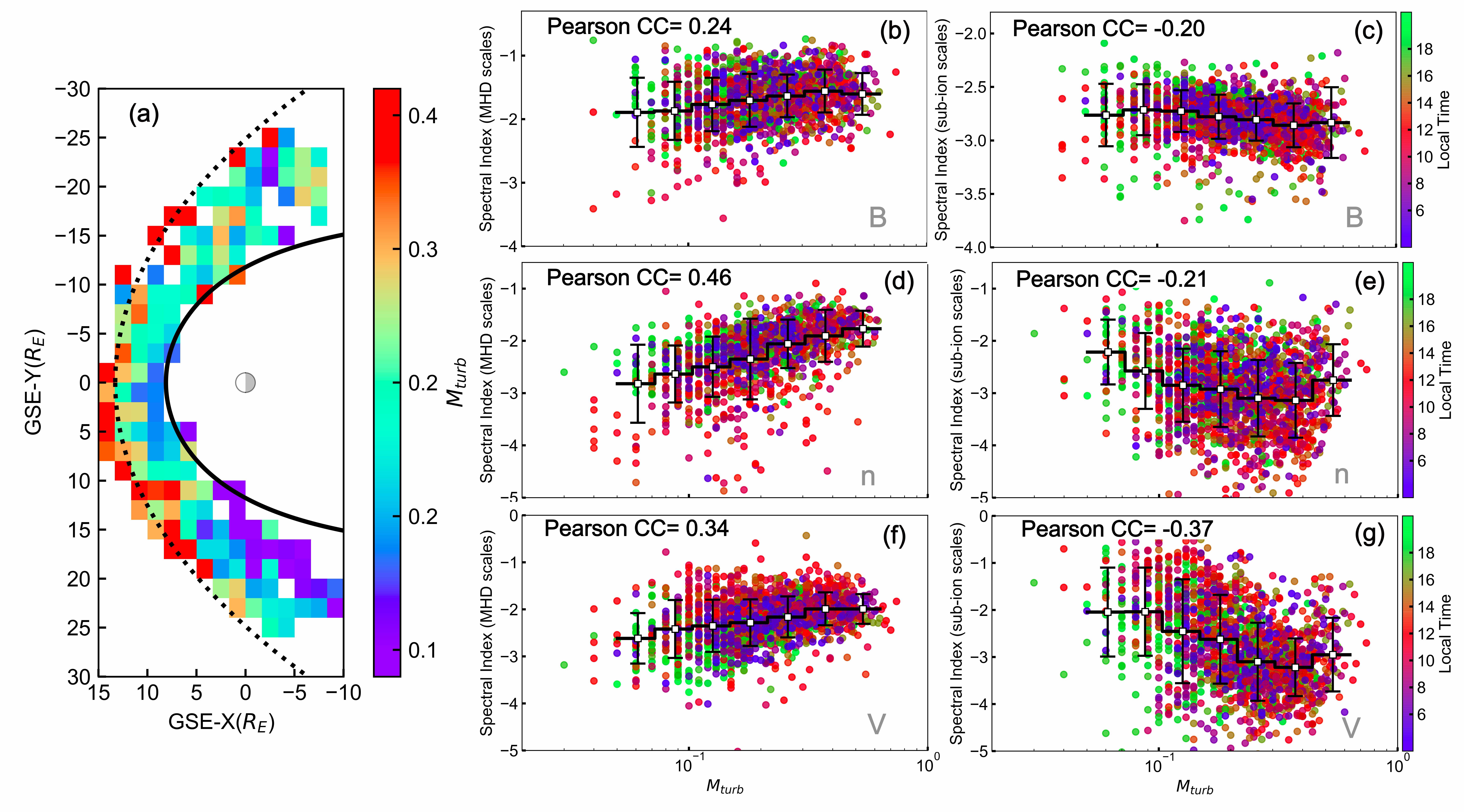}
\caption{Evolution of magnetosheath turbulences from the bow shock to the magnetopause. Panel (a) gives the 2D distribution of the turbulent sonic Mach number, $M_{\mathrm{turb}}$, in the GSE-XY plane of the magnetosheath. The black solid and the black dotted curves represent the nominal position of the Earth's magnetopause and the bow shock estimated from the models proposed by \citet{Shue97} and \citet{Chao02}, respectively. Panels (b), (d), (f) present the magnetic-field spectral indices, the density spectral indices, and the velocity spectral indices at MHD scales as a function of $M_{\mathrm{turb}}$. The results for sub-ion scales are accordingly given in panels (c), (e), (g). The color of data points indicates the local time. The black horizontal line represents the mean value for each bin, and the vertical line represents the standard deviation. CC represents the correlation coefficient between the spectral slope and $\log_{10} M_{\mathrm{turb}}$.}
\label{fig4}
\end{figure*}

Figure \ref{fig4} (a) gives the 2D distribution of the turbulent sonic Mach number, $M_{\mathrm{turb}}=\delta V/C_\mathrm{s}$ (where \textbf{$\delta V$} is the velocity fluctuation and $C_\mathrm{s}$ is the sound speed), in the GSE-XY plane of the magnetosheath. We find that $M_{\mathrm{turb}}$ is well organized from the bow shock to the magnetopause. Large $M_{\mathrm{turb}}$ ($>$ 0.4) are more likely to occur in the vicinity of the bow shock, and generally decreases to $<$ 0.1 closer to the magnetopause. From top to bottom, the middle three panels present the spectral indices of the magnetic-field, density, and velocity spectra at MHD scales as a function of $M_{\mathrm{turb}}$. The color of the date points indicates the local time (LT). The spectral indices at MHD-scales are all positively correlated with $M_{\mathrm{turb}}$, with CC = 0.24, 0.46, and 0.34, respectively. The mean spectral slops of magnetic-field, density and velocity spectra increase from -1.9, -2.8, -2.6 for $M_{\mathrm{turb}} < 0.08$ to -1.6, -1.8, -2.0 for $M_{\mathrm{turb}} > 0.6$, respectively. The right three panels accordingly give the results for sub-ion scales. Compared to MHD scales, we find negative correlations between the spectral indices and $M_{\mathrm{turb}}$ at sub-ion scales, with CC = -0.20, -0.21, and -0.37 for magnetic-field, density and velocity spectra, respectively. These are consistent with the results shown in Figure \ref{fig3}, especially in the quasi-parallel magnetosheath.

As shown in Figure \ref{fig1} (h) and Figure \ref{fig4} (b), there exist some $f^{-1}$-like magnetic-field spectra at MHD scales. \citet{Huang17} suggest that the random-like fluctuations generated behind the bow shock might lead to the formation of such $f^{-1}$-like magneti-field spectra. The steepening of magnetic-field spectral slope at MHD scales from the bow shock to the magnetopause is consistent with the picture that the turbulences downstream of the bow shock undergoes energy injection processes induced by, e.g. ion beams \citep{Lucek05}, forming shallower magnetic-field spectra at MHD scales, and then evolves to more fully-developed turbulences near the magnetopause. Early observations suggest that compressive turbulence in fast solar wind produces flat density spectra \citep{Marsch90} and show a good correlation between density fluctuations and sonic Mach number \citep{Klein93}. In the context of compressive MHD turbulence, small high-density compressed regions associated with slow and fast modes dominate the flattened spectrum \citep{LithGoldreich01}. In the supersonic regime, numerical simulations of hydrodynamic and MHD turbulence revealed that the density spectrum in the inertial range flattens when the root-mean-square sonic Mach number increases, while the velocity spectrum changes in the opposite sense \citep{Kritsuk07, KL07}. \citet{Banerjee13} attribut those results to the variability of the energy cascade rate due to compressibility. As shown in Figure \ref{fig4} (d), the density spectra at MHD scales flatten with increasing $M_\mathrm{turb}$, which is consistent with previous numerical and experimental studies in the supersonic regime \citep{Kritsuk07, KL07, Kowal07, Konstandin16, White19}. However, velocity spectral indices at MHD scales also flatten with the increasing of $M_{\mathrm{turb}}$, in contrast to the results in supersonic turbulence \citep{Kritsuk07, KL07}. Note that, $M_{\mathrm{turb}}$ is always less than 1 in the magnetosheath. 

\subsection{Dependence on upstream solar wind conditions}

\begin{table*}[!htbp]
\caption{Correlation coefficients between the turbulence spectral indices and $M_{\mathrm{A}}$, $M_{\mathrm{turb}}$ for different upstream solar wind conditions, such as the z-component of interplanetary magnetic field (IMF $B_\mathrm{z}$) and the solar wind speed ($V_\mathrm{p}$).}
\label{tab1}
\tablenum{1}
\centering
\begin{tabular}{ccccccccc}
\hline
\multicolumn{3}{c}{\multirow{2}{*}{}}   & \multicolumn{2}{c}{B} & \multicolumn{2}{c}{n} & \multicolumn{2}{c}{V} \\ \cline{4-9} 
\multicolumn{3}{c}{}                    &   MHD        &    sub-ion      & MHD   &    sub-ion       & MHD    &  sub-ion         \\ \hline
\multirow{4}{*}{ IMF $B_\mathrm{z}$} & \multirow{2}{*}{$B_\mathrm{z} > 0.5$ nT} & $M_{\mathrm{A}}$ & 0.40  & 0.16    & 0.04 &0.08  &-0.25 & 0.04    \\ \cline{3-9} 
                  &                   & $M_{\mathrm{turb}}$ & 0.31    &-0.12     &0.53    &-0.27   &0.39     &-0.38     \\ \cline{2-9} 
                  & \multirow{2}{*}{$B_\mathrm{z} < -0.5$ nT} & $M_{\mathrm{A}}$ & 0.39  &  -0.04   &  0.09  & 0.08   &-0.24     & 0.00    \\ \cline{3-9} 
                  &                   & $M_{\mathrm{turb}}$ &  0.26     &  -0.27  &0.57    & -0.09   &0.25     &-0.37      \\ \hline
\multirow{4}{*}{Flow speed $V_\mathrm{p}$} & \multirow{2}{*}{$V_\mathrm{p} > 400$ km/s} & $M_{\mathrm{A}}$ &  0.41& 0.03    & 0.06   & 0.07  & -0.21  &-0.01     \\ \cline{3-9} 
                  &                   &  $M_{\mathrm{turb}}$& 0.26  &-0.12  &  0.50  &  -0.16  &0.32    & -0.42   \\ \cline{2-9} 
                  & \multirow{2}{*}{$V_\mathrm{p} < 400$ km/s} & $M_{\mathrm{A}}$ &  0.37 & 0.02  &0.06    &0.07    & -0.27  & 0.00   \\ \cline{3-9} 
                  &                   &  $M_{\mathrm{turb}}$&0.24     &  -0.26 & 0.49   & -0.17   & 0.28  &-0.33    \\ \hline
\end{tabular}
\end{table*}

It is often believed that the upstream solar wind conditions, e.g., the angle $\Theta_{\mathrm{Bn}}$, flow speed $V_\mathrm{p}$ and z-component of the interplanetary magnetic filed $B_\mathrm{z}$, modify the structure, dynamics, and dissipation processes in the magnetosheath \citep{Dimmock14}. To analyze the influence of the upstream solar wind conditions on the magnetosheath turbulence properties, we compare our results of IMF $B_\mathrm{z} > 0.5$ nT vs. IMF $B_\mathrm{z} < -0.5$ nT, $V_\mathrm{p} > 400$ km/s vs. $V_\mathrm{p} < 400$ km/s in Table \ref{tab1}. In general, we find no significant discrepancies of the correlation coefficients between the three turbulence spectral indices (magnetic-field, density and velocity spectra) and $M_{\mathrm{A}}$ or $M_{\mathrm{turb}}$ under different upstream solar wind conditions. In addition，we do not find any systematical differences of the spectral indices as a function of $M_{\mathrm{A}}$ and $M_{\mathrm{turb}}$ either, which is not shown here. These results indicate that the spatial evolution of magnetosheath turbulence is largely independent of the upstream solar wind conditions.

\section{Summary}

The Earth's magnetosheath is a highly turbulent region bounded by the bow shock and the magnetopause, within which the solar wind reduces from supersonic to subsonic flow speeds after crossing the bow shock, and returns to a supersonic flow in the flanks. Thus, the magnetosheath provides a good natural laboratory to investigate the spatial evolution of space plasma turbulence from a small Alfv\'en Mach number, $M_\mathrm{A}$, to a large $M_\mathrm{A}$. By means of simultaneous observations of magnetic field and plasma moments with unprecedented high time resolution provided by the MMS mission, using 1841 burst-mode segments of MMS-1 from 2015/09 to 2019/06, we study two aspects of the spatial evolution of magnetosheath turbulence at both MHD and sub-ion scales statistically. 

From the sub-solar region to the flanks, $M_{\mathrm{A}}$ increases. The spectral index of magnetic-field spectra at MHD scales changes from -2 for $M_{\mathrm{A}} < 0.8$ to -3/2 for $M_{\mathrm{A}} > 5$, presenting a positive correlation with $M_{\mathrm{A}}$. For the density spectra at MHD scales, the spectral index remains $\sim$ -7/3 for different $M_{\mathrm{A}}$. The spectral index of velocity spectra at MHD scales changes from -2 for $M_{\mathrm{A}} < 0.8$ to -5/2 for $M_{\mathrm{A}} > 5$, presenting a negative correlation with $M_{\mathrm{A}}$. At sub-ion scales, we find no obvious relations between the spectral index and $M_{\mathrm{A}}$ for all the three types of spectra.

We also investigate the evolution of magnetosheath turbulence from the bow shock to the magnetopause. We use two parameters to represent the relative distance of the spacecraft from the bow shock or the magnetopause. One is the widely used fractional distance, $D_{\mathrm{frac}}$. The other is the turbulent sonic Mach number, $M_{\mathrm{turb}}$, which generally decreases from $>$ 0.4 in the bow shock vicinity to $<$ 0.1 near the magnetopause. At MHD scales, the magnetic-field, density and velocity spectra steepen from the bow shock to the magnetopause, especially in the quasi-parallel magnetosheath. The mean value of spectral index for magnetic-field spectra changes from -1.46 in the vicinity of the bow shock to -1.94 in the vicinity of the magnetopause. For density and velocity spectra, the index changes from -1.87 to -2.44 and from -1.92 to -2.18, respectively. At sub-ion scales, in-distinctive opposite trends of radial evolution are observed. All three types of spectra present a slight flattening from the bow shock to the magnetopause. Furthermore, the spectral indices are positively correlated with $M_\mathrm{turb}$ at MHD scales, and negatively correlated with $M_\mathrm{turb}$ at sub-ion scales. The break frequency of magnetic-field and velocity spectra increases when approaching the magnetopause, from 0.24 to 0.33 Hz and from 0.32 to 0.43 Hz, respectively. For the density spectra, the break frequency remains around 0.3 Hz. Similar results, except for some minor differences, are found in the quasi-perpendicular magnetosheath. Our results might expand our knowledge on sub-sonic compressive MHD turbulence in the magnetosheath and contribute to understand the transition from the sub-sonic to the supersonic regime.

We also discuss the influences of upstreams solar wind conditions, e.g., northward IMF ($B_\mathrm{z} >$ 0.5 nT) vs. southward IMF ($B_\mathrm{z} <$ -0.5 nT), fast wind ($V_\mathrm{p} > 400$ km/s) vs. slow wind ($V_\mathrm{p} < 400$ km/s), on the spatial evolution of magnetosheath turbulence. We find no significant dependencies on the upstream parameters, suggesting that the spatial evolution of magnetosheath turbulence is largely independent of the upstream solar wind conditions. 

\acknowledgments
The authors thank the MMS team for providing the FGM and FPI data at the MMS Science Data Center (\url{https://lasp.colorado.edu/mms/sdc/public}), and NASA CDAWEB (\url{https://cdaweb.sci.gsfc.nasa.gov/index.html/}) for providing the OMNI data. Special thanks to Dr. B. B. Tang, Dr. W. Y. Li, and Prof. X. C. Guo for helpful discussions. This work was supported by Strategic Priority Research Program of Chinese Academy of Sciences grant No. XDA17010301), NNSFC grants 41874203, 41574169, 41574159, 41731070, Young Elite Scientists Sponsorship Program by CAST, 2016QNRC001, and grants from Chinese Academy of Sciences (QYZDJ-SSW-JSC028, XDA15052500). H. Li was also supported by Youth Innovation Promotion Association of the Chinese Academy of Sciences and in part by the Specialized Research Fund for State Key Laboratories of China. D.V. is supported by the STFC Ernest Rutherford Fellowship ST/P003826/1 and STFC Consolidated Grant ST/S000240/1.

\end{document}